\documentclass[11pt,a4paper]{article}
\pdfoutput=1
\usepackage{jheppub}
\usepackage[page,toc,titletoc,title]{appendix}
\usepackage{mathtools}
\usepackage[all]{hypcap}
\usepackage{xcolor}
\usepackage{url}
\usepackage[utf8]{inputenc}
\allowdisplaybreaks%
\begin{document}


{\raggedleft CTPU-PTC-20-26, PNUTP-20-A11\\}

\title{
  Light Higgsinos for electroweak naturalness \\
  in mirage-mediated high-scale supersymmetry
}

\affiliation[a]{Department of Physics, Pusan National University, Busan 46241,
  South Korea}
\affiliation[b]{Center for Theoretical Physics of the Universe,\\
  Institute for Basic Science (IBS), Daejeon 34051, South Korea}

\author[a]{Kwang~Sik~Jeong}
\author[b]{and Chan~Beom~Park}

\emailAdd{ksjeong@pusan.ac.kr}
\emailAdd{cbpark@ibs.re.kr}

\abstract{
Mirage mediation realized in the  Kachru-Kallosh-Linde-Trivedi (KKLT) flux compactification can
naturally suppress the up-type Higgs soft mass at low energy scales.
As a result,
compared to the conventional scenarios,
the degree of electroweak fine-tuning can be reduced
further up to by
a loop factor if the Higgsinos are much lighter than the heavy
Higgs doublet.
Interestingly,  this feature holds even in high-scale supersymmetry
as long as the gauge coupling unification,  which is required as a prerequisite
for mirage mediation,  accommodates such light Higgsinos.
Under the experimental constraints on the observed Higgs boson,
it turns out that mirage mediation can exhibit low electroweak fine-tuning better than a few
percent for stops between about $2$ and $6$~TeV,  i.e.,  at the same
level as in the weak scale supersymmetry,  if the Higgsinos are around
or below a few hundred GeV.
}

\maketitle

\section{Introduction}

Supersymmetry (SUSY) is a unique extension of the Poincar\'e
spacetime symmetry,  and it has been extensively explored as one of the
most plausible candidates for physics beyond the Standard Model (SM)
because it can address various problems of the SM such as dark matter,
unification,  and cosmological inflation~\cite{Nilles:1983ge, Haber:1984rc}.
In particular,  it has been strongly motivated by the hope to explain
the huge hierarchy between the weak scale and the unification or Planck scale.
However,  the current experimental results indicate that SUSY,
if realized in nature,  should be above the TeV scale.
In this circumstance, as it becomes more difficult to stabilize
the electroweak scale against large radiative corrections from unknown
ultraviolet physics, one might need to rely on some other mechanism,
such as the cosmological relaxation~\cite{Graham:2015cka},
in order to explain the hierarchy between the SUSY breaking scale and
the weak scale.

If SUSY exists above the TeV scale,  an important question worth asking
is whether it can still solely solve the gauge hierarchy problem,
which has been believed to be one of the important virtues of SUSY\@.
In this paper,  we explore the possibility of relaxing the electroweak
fine-tuning in high-scale SUSY from mirage
mediation~\cite{Choi:2005uz}
that is realized in the KKLT flux compactification~\cite{Kachru:2003aw, Choi:2005ge}.
It has been noticed that TeV scale mirage mediation can naturally
suppress the low-energy value of the up-type Higgs soft mass through
the combined effect of anomaly and moduli mediations~\cite{Choi:2005hd, Choi:2006xb}.
Consequently, it can reduce the degree of fine-tuning
for electroweak symmetry breaking roughly by a loop factor
if the Higgsinos are much lighter than the heavy Higgs doublet.
We stress that this naturalness feature persists even in mirage-mediated high-scale
SUSY as long as the gauge coupling unification,  which is
a prerequisite for mirage mediation,  allows the Higgsinos to be very light compared
to other sparticles.

Mirage mediation,  in which anomaly mediation is comparable to moduli
mediation in strength,  effectively corresponds to pure moduli
mediation transmitted at the mirage messenger scale, where there is no
physical threshold.
The electroweak fine-tuning can considerably be reduced when the
mirage messenger-scale exists around the SUSY breaking scale.
It only requires a proper choice of the discrete numbers associated with
the string moduli sector~\cite{Choi:2005hd, Choi:2006xb}.
We find that such a mirage-mediation scheme should make stops
heavier than about $2$~TeV in order to give the correct mass to the
SM-like Higgs boson, but nonetheless, it can exhibit low electroweak
fine-tuning better than a few percent for stops below $6$~TeV if
the Higgsinos are around or below a few hundred GeV.
One of the important consequences of mirage mediation is that it leads
to highly compressed spectra of gauginos and sfermions, thus allowing us
to precisely fix the Higgsino mass by the condition of gauge coupling
unification within the minimal-supersymmetric SM (MSSM).

This paper is organized as follows.
In Sec.~\ref{sec:higgs}, we present a review on how the Higgs
sector parameters are generated in the mirage mediation scheme and
then discuss the way to reduce the degree of electroweak fine-tuning
in high-scale SUSY while making the SM-like Higgs boson compatible
with the current experimental data.
The viable parameter region leading to natural electroweak symmetry
breaking is examined in Sec.~\ref{sec:naturalness}.
The final section is for conclusions.

\section{\label{sec:higgs}Higgs Sector in Mirage Mediation}

In the MSSM,  the $Z$-boson mass is determined by the minimization
condition of the Higgs scalar potential,
\begin{equation}
\label{min-condition}
\frac{m^2_Z}{2} =
-|\mu|^2 + \frac{m^2_{H_d} - m^2_{H_u}\tan^2\beta } {\tan^2\beta -1},
\end{equation}
where the involved parameters in the expression above are evaluated
around the weak scale,  and $\tan\beta$ is the ratio of the vacuum
expectation values (VEVs) of the Higgs doublets,  $H_u$ and $H_d$.
The $Z$-boson mass is quite sensitive to the variation of $m^2_{H_u}$
for moderate to large values of $\tan\beta$,  and in general,  the
electroweak fine-tuning becomes more severe as the SUSY breaking scale
increases.
Furthermore,  the successful electroweak symmetry breaking also
requires a sizable Higgs mixing term,
\begin{equation}
|B\mu| =  \frac{\sin2\beta}{2} (m^2_{H_d} + m^2_{H_u} + 2|\mu|^2)
\end{equation}
at the weak scale.
The degree of fine-tuning quantifies to what extent the $Z$-boson mass
is sensitive to the variations of the Higgs sector parameters.
As shall be discussed later,  the correct estimation of fine-tuning requires
the addition of  the loop potential  to the
renormalization group (RG) improved tree-level potential.

Combined with the nonobservation of sparticles around the TeV scale
in collider experiments so far,  the measured Higgs boson mass
$m_h\simeq 125$~GeV indicates that SUSY may show up around or above the
multi-TeV scale if realized in nature.
It is thus generally expected that the electroweak symmetry breaking
would require severe fine-tuning, at the level of $0.1$\%,
unless some other mechanism to cure the fine-tuning is invoked.
Having said that,  if the Higgsinos are relatively light as compared to
other sparticles,  the degree of fine tuning can be reduced by a
sizable amount in a mediation scheme such that the up-type Higgs
soft mass is naturally suppressed at low-energy scales,
\begin{equation}
|m^2_{H_u}|
\lesssim
|\mu|^2
\ll m^2_{H_d}.
\end{equation}
In terms of the electroweak fine-tuning,  the stop sector is
particularly important since it affects the RG
running of the up-type Higgs soft mass via the top-quark Yukawa
coupling.
The stop radiative contribution to $m_{H_u}^2$ is given by
\begin{equation}
\label{mhu}
\delta m^2_{H_u} \simeq -\frac{3y^2_t}{4\pi^2}
m^2_{\tilde t} \ln\left( \frac{\Lambda_{\rm mess}}{m_{\tilde t} } \right),
\end{equation}
where $m_{\tilde t}$ is the stop mass,  and $\Lambda_{\rm mess}$ denotes
the messenger scale
at which SUSY breaking is transmitted to the MSSM sector.
It shows that having a suppressed value of $m^2_{H_u}$ below the order
of $m^2_{\tilde t}$ at low energy is unattainable unless the messenger
scale is very low.
In this regard,  mirage-mediated SUSY breaking,  realized in the KKLT
flux compactification,  is of particular interest because it
effectively corresponds to pure moduli mediation transmitted at the
mirage messenger scale,  $M_{\rm mir}$,  while not being bothered by
physical thresholds at the scale.
The mirage messenger scale can be very low,  depending on the relative
strength of anomaly mediation.

It should be noted that the inclusion of the loop potential effectively amounts to the shift,
roughly given by
\begin{equation}
\label{mhu2}
m^2_{H_u} \to m^2_{H_u} -\frac{3y^2_t}{8\pi^2} m^2_{\tilde t},
\end{equation}
in the RG-improved tree-level potential.
The above correction is quadratically sensitive to the
sparticle mass scale
and remains unsuppressed in mirage mediation as is the case in other
scenarios.
The stop contribution~(\ref{mhu}),  which generally makes the Higgs mass
more sensitive to UV scales  due to
the logarithmic factor, is the one we wish to suppress within the KKLT framework.
The minimization condition~(\ref{min-condition}) requires
the Higgsino mass parameter roughly to be
\begin{equation}
|\mu|^2 \approx \frac{m^2_{H_d}}{\tan^2\beta} - m^2_{H_u},
\end{equation}
for moderate or large $\tan\beta$.
Moreover,
the gauge coupling unification is required for the
sparticles to exhibit the mirage pattern,  and it can be
achieved in high-scale SUSY as precisely as in weak-scale SUSY if the
Higgsinos are light,  as discussed below.

In the KKLT flux compactification, sparticle masses exhibit the mirage
mediation pattern as a result of mixed anomaly and moduli mediation.
Provided that the gauge couplings unify at $M_{\rm GUT}\sim 10^{16}$~GeV,
the gaugino masses are given by
\begin{equation}
M_a(Q) = M_0 \left\{
1 - \frac{b_a g^2_a }{4\pi^2} \ln\left( \frac{M_{\rm mir}}{Q} \right)
\right\},
\end{equation}
at the renormalization scale $Q$,
with $b_a$ being the coefficients of one-loop beta functions.
Here the relative strength between anomaly and moduli mediation is
measured by the parameter defined as
\begin{equation}
  \alpha  \equiv  \frac{m_{3/2}}{M_0 \ln(M_{Pl}/m_{3/2})},
\label{eq:alpha}
\end{equation}
with $m_{3/2}$ being the gravitino mass.
Then,  the mirage messenger scale is determined by
\begin{equation}
M_{\rm mir} = M_{\rm GUT} \left( \frac{m_{3/2}}{M_{Pl}}
\right)^{\alpha/2}.
\label{eq:M_mir}
\end{equation}
In the KKLT setup,  the $\alpha$ parameter has a positive rational
number of order unity at leading order.
%
The soft SUSY breaking parameters of scalar fields also take the mirage
pattern
\begin{eqnarray}
A_{ijk}(Q) &=& M_0 \left\{
1  + \frac{\gamma_i + \gamma_j + \gamma_k}{8\pi^2}
\ln \left( \frac{M_{\rm mir} }{Q} \right)
\right\},
\nonumber \\
m^2_i(Q) &=& M^2_0 \left\{
c_i + \frac{1}{4\pi^2} \left(
\gamma_i
- \frac{\dot \gamma_i}{2}
 \ln \left( \frac{M_{\rm mir} }{Q} \right)
\right)  \ln \left( \frac{M_{\rm mir} }{Q} \right)
\right\},
\end{eqnarray}
for a proper choice of matter modular weights,
where $\gamma_i$ is the anomalous dimension of the corresponding
field,  and the dot denotes differentiation with respect to $Q$.
Here $c_i$ parametrizes the moduli-mediated contribution, given by
\begin{equation}
c_i = 1- n_i + {\cal O}\left( \frac{1}{8\pi^2} \right),
\end{equation}
including quantum corrections from string loops and higher-order
$\alpha^\prime$ corrections.
The modular weight $n_i$ has a rational number of order unity, depending on
the location of the matter in extra dimensions.
The above mirage pattern of soft scalar parameters requires the
matter fields having a Yukawa coupling $y_{ijk}$ to have modular
weights satisfying
\begin{equation}
\label{mirage-condition}
 n_i + n_j + n_k = 2,
\end{equation}
unless $y^2_{ijk}$ is smaller than $1/8\pi^2$.

In the generalized KKLT scenario with dilaton-moduli mixing in gauge
kinetic functions,  the $\alpha$ parameter can have various values,
depending on discrete numbers associated with dilaton and moduli
couplings~\cite{Choi:2005hd, Choi:2006xb}.
Out of this framework,  we can have $\alpha = 2$ while allowing an
extra-dimensional interpretation for the origin of SUSY breaking.
The scenario of mirage mediation with $\alpha=2$ and $n_{H_u}=1$ is of
particular interest because it can considerably reduce the electroweak
fine-tuning.
For $\alpha=2$,  the mirage messenger scale is fixed at
\begin{equation}
  M_{\rm mir} = \frac{M_{\rm GUT}}{M_{Pl}} m_{3/2} \sim M_0,
\end{equation}
and thereby is around the SUSY breaking scale.
Here we have used the relation that the ratio between the Planck and
unification scales is numerically close to a loop factor.
Besides, if $n_{H_u}=1$,  the up-type Higgs soft mass at $M_{\rm mir}$
is loop-suppressed relative to the squared soft scalar mass of other
scalar fields.
Interestingly,   the electroweak fine-tuning remains tempered even for
$M_0$ being the multi-TeV energy regime as long as the gauge coupling
unification is maintained.
As a benchmark for accomplishing a natural electroweak symmetry breaking,
let us assign matter modular weights as
\begin{equation}
\label{ni}
n_{H_u} = 1, \quad n_{H_d} = 0,
\quad
n_{\rm quarks} = n_{\rm leptons} = \frac{1}{2},
\end{equation}
where $n_{H_u}=1$ is required to reduce the electroweak fine-tuning,
while there is no strict requirement on the  modular weights of other fields
as long as the mirage conditions~(\ref{mirage-condition}) are satisfied.\footnote{
  A matter field has $n_i=0$,  $1/2$,  or $1$ depending on its location
  in extra dimensions. The modular weight can have a different
  rational number in a generalized KKLT with an anomalous U$(1)$ gauge
  symmetry~\cite{Choi:2006bh}.
  One may assign different modular weights to the down-type Higgs and
  matter fields while keeping the flavor universality and satisfying
  the mirage condition.
  It can change the degree of electroweak fine-tuning only slightly if
  the modular weights are of order unity.
  Still, we need $n_i<1$ for quarks and leptons since otherwise there
  would appear dangerous color or charge-breaking minima.
  See Ref.~\cite{Kawamura:2017qey} for a recent study of other
  viable assignment for matter modular weights.
}
For the above choice of modular weights,   the mirage conditions are satisfied
with good accuracy in the region with $\tan^2\beta \ll (m_t/m_b)^2$
because the effects of Yukawa couplings (other
than that of top quark) can be neglected in the RG evolution of scalar
soft parameters.
Here we have assumed that the sfermions are supposed to feel SUSY
breaking through flavor-conserving interactions unless they are
heavier than about $100$~TeV.
Note that mirage mediation preserves CP symmetry as a result of
axionic shift symmetries associated with the moduli.
One can then find that the gaugino masses are universal
\begin{eqnarray}
\label{soft1}
M_a &=& M_0,
\end{eqnarray}
and the soft SUSY breaking parameters for scalar fields read
\begin{eqnarray}
\label{soft2}
A_t &\simeq& M_0,
\nonumber \\
A_b &\simeq& A_\tau \simeq 2M_0,
\nonumber \\
m^2_{H_u} &=&
{\cal O}\left( \frac{M^2_0}{8\pi^2} \right),
\nonumber \\
m^2_{H_d} &\simeq&
M^2_0,
\nonumber \\
m^2_{\tilde q} &\simeq& m^2_{\tilde \ell}
\simeq
\frac{M^2_0}{2},
\end{eqnarray}
at the energy scale $Q=M_{\rm mir}\sim M_0$.
Here $\tilde q$ and $\tilde \ell$ refer to squarks and sleptons,  respectively,
dropping the flavor indices.

Let us now examine how light the Higgsino can be in the mirage mediation scheme.
In the MSSM, the gauge coupling unification can successfully be attained
in the absence of heavy threshold corrections
if the sparticle spectrum satisfies the following
relation~\cite{Krippendorf:2013dqa,Jeong:2017hgz}:\footnote{
If high-scale threshold corrections are sizable,  the gauge coupling
unification can accommodate a much wider spectrum of sparticle
masses~\cite{Ellis:2017erg},
allowing for lighter Higgsinos than required by the relation~(\ref{mu-gut}).
In such a case, however, the mirage-mediated pattern of sparticle masses is spoiled,
making it difficult to suppress the up-type Higgs soft mass at low-energy scales.
See Appendix~\ref{sec:threshold_corr} for details.
%
}
\begin{equation}
  \label{mu-gut}
  \frac{|\mu|}{m_\ast}
  \left( \frac{m_H}{m_\ast} \right)^{1/4}
  \left( \frac{M_2}{m_\ast} \right)^{1/3}
  \left( \frac{M_2}{M_3} \right)^{7/3}
  \left( \frac{m_{\tilde{\ell}}}{m_{\tilde q}} \right)^{1/4}
  = 1,
\end{equation}
with $m_H$ being the heavy Higgs doublet mass.
The scale parameter $m_\ast$ determines the precision of the gauge coupling unification.
It should be above a few hundred GeV and below $10$~TeV to achieve the
unification within a deviation of a few percent.
As a consequence of the relation,  for mirage mediation with $\alpha=2$
and the modular weights given by the relation~(\ref{ni}),
we find the Higgsino mass as
\begin{equation}
\label{unification}
|\mu| \simeq 130~{\rm GeV}
\left( \frac{m_\ast}{0.5~{\rm TeV}} \right)^{19/12}
\left( \frac{M_0}{5~{\rm TeV} } \right)^{-7/12},
\end{equation}
assuring the gauge coupling unification at $M_{\rm GUT}\sim 10^{16}$~GeV.
Therefore,  mirage mediation can be realized in high-scale SUSY while
accommodating light Higgsinos,  even around the weak scale.
In high-scale SUSY with light Higgsinos,  the lightest neutralinos and
the lightest chargino are dominated by the Higgsino components,  and
their mass difference is estimated by
\begin{equation}
\Delta m = m_{\chi^+_1} - m_{\chi^0_1}
\simeq
0.8\,{\rm GeV} \left( \frac{M_0}{5~{\rm TeV}} \right)^{-1}
+ 0.3\,{\rm GeV}\left( \frac{|\mu|}{300~{\rm GeV}}\right)^{0.15},
\end{equation}
including the contributions from gauge boson loops~\cite{Thomas:1998wy}.
Here we have used the sparticle mass pattern of mirage mediation with
$\alpha = 2$,  where the gauginos have a degenerate mass spectrum,
$M_a\simeq M_0$.
Although it would be difficult to detect the lightest chargino at the LHC
because of the degenerate mass spectrum,
we expect that future lepton colliders may probe the signals
from the processes of $e^+e^-\to \chi^0_1 \chi^0_2 \gamma$ or
$\chi^+_1 \chi^-_1 \gamma$,  mediated by a virtual $Z$ boson or
photon~\cite{Baer:2011ec, Berggren:2013vfa}.

We close this section by discussing the dynamical generation of the
Higgs mixing parameter.
In models with sizable anomaly mediation,  the Higgs mixing parameter
is generally of the order of the gravitino mass, which is too large
to induce electroweak symmetry breaking correctly.
To generate it at the right scale,  i.e.,~to have $B\sim M_0$,  one
can extend the Higgs sector, for instance,
by coupling it to the K\"ahler modulus $T$ through the nonperturbative
superpotential term from hidden gaugino condensation
\begin{equation}
\Delta W = A e^{-\frac{1}{2} a T} H_u H_d,
\end{equation}
for $T$ stabilized by the nonperturbative superpotential term,
$\Delta W_{\rm np} = A_0 e^{-a T}$,
in the KKLT setup~\cite{Choi:2005hd, Choi:2006xb}.
Another interesting way to obtain $B\sim M_0$ is to consider the effective K\"ahler potential
\begin{equation}
\Delta K = \kappa \frac{\bar S}{S} H_u H_d + {\rm H.c.},
\end{equation}
in the model,  where the singlet scalar $S$ is radiatively
stabilized~\cite{Nakamura:2008ey,Choi:2009qd}.
In this case, the phase component of $S$ can play the role of the
axion that can provide a solution to the strong CP problem.
It is also worthwhile to note that the fermionic partner,  the axino,
can contribute to the dark matter of the Universe while
avoiding the cosmological problems arising when the Universe experiences
the modulus-dominated phase~\cite{Nakamura:2008ey}.

\section{\label{sec:naturalness}Electroweak  Naturalness}

In this section we examine how naturally the electroweak symmetry
breaking arises in the mirage mediation with $\alpha=2$ and $n_{H_u}=1$.
In this benchmark scenario,  the Higgs soft masses are
given by
\begin{equation}
|m^2_{H_u}| \sim \frac{m^2_{\tilde t}}{8\pi^2}
\,\ll\,
m^2_{H_d} \sim m^2_{\tilde t},
\end{equation}
at the SUSY breaking scale,
with the stop mass $m_{\tilde t}\simeq M_0/\sqrt 2$ as presented in the relation~(\ref{soft2}).
The value of $m^2_{H_u}$ is rather sensitive to the renormalization scale at energy scales
around the mirage messenger scale,  which is close to the stop mass $m_{\tilde t}$.
To cancel the dependence of the Higgs sector parameters on the renormalization scale,
one needs to include one-loop effective potential as
\begin{equation}
V = V_{\rm tree} + \Delta V,
\end{equation}
where $V_{\rm tree}$ is the RG improved tree-level potential, and the loop potential
$\Delta V$ is generated dominantly by the loops involving
third-generation sfermions.
In the electroweak symmetry breaking conditions derived from $V_{\rm tree}$,
the inclusion of $\Delta V$ effectively corresponds to the replacement
\begin{equation}
m^2_i \to m^2_i + t_i,
\end{equation}
for $i \in \{H_u$, $H_d\}$.
Here the tadpoles $t_i$ are calculable from the squared mass matrix after electroweak
symmetry breaking~\cite{Baer:2012cf}, and have the values of
\begin{equation}
|t_i| \sim \frac{M^2_0}{8\pi^2} .
\end{equation}
The above radiative corrections represent the quadratic sensitivity of the Higgs mass
to the scale of sparticle masses.
Our scenario is to suppress the RG evolved up-type Higgs mass squared down to the
order of the tadpoles,
which is naturally achieved by taking $\alpha=2$ and $n_{H_u}=1$ in mirage mediation.
As a result,  the degree of electroweak fine-tuning can be reduced further,
up to a loop factor, compared to the conventional scenarios.

For the correct estimation of the degree of electroweak fine-tuning, one should note that
the Higgs mass parameters are given by  $m^2_i = c_i M^2_0$
at the mirage messenger scale, in which $c_i$
generally does receive model-dependent
quantum corrections from string loops and
higher-order $\alpha^\prime$ corrections.
This implies that the RG improved Higgs mass parameters include a correction of the order
of $M^2_0/8\pi^2$,
thus it is as sizable as the tadpoles $t_i$ from the loop potential.
Taking this into account,
we examine the electroweak symmetry breaking using $V_{\rm tree}$ by
replacing the Higgs mass parameters as
\begin{equation}
m^2_i \to
\tilde m^2_i = \left( 1 - n_i + \frac{\kappa_i}{8\pi^2} \right) M^2_0,
\end{equation}
for constants $\kappa_i$ of order unity or below.
The parameters $\kappa_i$ encompass all the renormalization
scale dependence and higher-order effects;
the tadpole $t_i$,  higher-order moduli mediated contributions,  and
model-dependent stringy higher-order corrections.
It is obvious that $\kappa_i$ should be treated as a free parameter of order unity or below
due to the stringy corrections.
We emphasize that setting $\kappa_{H_u}$ to be a free
parameter is quite important in our scenario because $n_{H_u}=1$
gives $\tilde m^2_{H_u} = \kappa_{H_u} M^2_0 / 8\pi^2$.
In our analysis below, the value of $\kappa_{H_u}$ is fixed by
imposing the minimization condition of the Higgs potential.

Let us now explore the parameter region leading to the correct electroweak symmetry
breaking while satisfying the current experimental constraints.
The minimization conditions now read
\begin{eqnarray}
\label{Hmin-condition}
\frac{1}{2}m^2_Z &=& -|\mu|^2
+ \frac{\tilde m^2_{H_d} - \tilde m^2_{H_u} \tan^2\beta}{\tan^2\beta -1},
\nonumber \\
\sin2\beta &=&
\frac{2|B\mu|}{\tilde m^2_{H_d} +\tilde m^2_{H_u} + 2 |\mu|^2},
\end{eqnarray}
with the Higgs mass parameters given by
\begin{eqnarray}
\tilde m^2_{H_u} &=& \frac{\kappa_{H_u}}{8\pi^2}M^2_0 ,
\nonumber \\
\tilde m^2_{H_d} &=&   \left( 1 + \frac{\kappa_{H_d}}{8\pi^2} \right)M^2_0.
\end{eqnarray}
Here the Higgs sector parameters must satisfy
\begin{eqnarray}
  \tilde{m}_{H_d}^2 + \tilde{m}_{H_u}^2 + 2 |\mu|^2
  & > 2 |B \mu|,
  \nonumber \\
  \left( \tilde{m}_{H_d}^2 + |\mu|^2 \right)
  \left( \tilde{m}_{H_u}^2 + |\mu|^2 \right)
  & < |B \mu|^2,
\end{eqnarray}
for the scalar potential to be bounded from below and to have a minimum at
nonzero Higgs VEVs.
The minimization conditions are  insensitive to
the value of $\kappa_{H_d}$,  so we will simply set $\kappa_{H_d}=0$
in the numerical analysis.
The viable parameter region can then be examined by scanning over the
two-dimensional space of
\begin{equation}
\{ M_0,\,\,\tan\beta \} .
\end{equation}
The value of $\mu$ is fixed by the unification condition~(\ref{unification})
for a given value of $m_\ast$,  and the $B$ and $\kappa_{H_u}$ values
are obtained by the minimization conditions~(\ref{Hmin-condition}).

At present, one of the most important constraints on models with an
extended Higgs sector is from the measurement of the Higgs boson mass.
The viable parameter region of $\{M_0$, $\tan\beta\}$ can thus be
found by requiring the SM-like Higgs boson to have $m_h \simeq 125$~GeV.
In the MSSM,  the mass of the lightest neutral Higgs boson reads
\begin{equation}
m^2_h|_{\rm tree} = \frac{1}{2}
\left( m^2_Z + m^2_A - \sqrt{
(m^2_Z + m^2_A)^2 - 4m^2_Z m^2_A \cos^2 2\beta
} \right)
\simeq m^2_Z \cos^2 2\beta,
\end{equation}
at tree level.
The last approximation holds in the decoupling limit with $m_A \gg m_Z$.
Higher-order corrections to the Higgs boson mass arise mainly via the
loops of third-generation sfermions,  and the gluinos also take part in
at two-loop level.
Using the effective field theory approach with RG improvements,
one finds the Higgs boson mass in the MSSM to be
\begin{equation}
\label{Higgs-mass}
m^2_ h = m^2_h|_{\rm tree}
+ \frac{3r }{4\pi^2} \frac{\bar m^4_t}{v^2},
\end{equation}
where $r$ is given by
\begin{eqnarray}
r &=& t + \frac{X^2_t}{M^2_S} \left( 1 - \frac{X^2_t}{12 M^2_S} \right)
+ \left( \frac{4\alpha_s}{3\pi} - \frac{5\bar m^2_t}{16\pi^2 v^2}\right) t
\nonumber \\
&&
+\,
 \frac{1}{16\pi^2} \left( \frac{3}{2} \frac{\bar m^2_t}{v^2} -32 \pi \alpha_s \right)
 \left\{
 \frac{X^2_t}{M^2_S}  \left( 2 - \frac{X^2_t}{6M^2_S} t +t^2 \right)
 \right\},
\end{eqnarray}
up to two-loop leading corrections~\cite{Carena:1995wu, Haber:1996fp,Carena:2000dp},
with $v\simeq 174$~GeV and $t\equiv \ln(M^2_S/\bar m^2_t)$.
Here $M^2_S$ is the average of the squared masses of two stops,
and $X_t$ is the stop mixing parameter defined by
\begin{equation}
X_t = A_t - \frac{\mu}{\tan\beta},
\end{equation}
with $A_t$ being the stop trilinear coupling.
In the mirage mediation under consideration,  the stop sector has
\begin{equation}
A_t \simeq M_0, \quad M_S \simeq \frac{M_0}{\sqrt 2},
\end{equation}
because the modular weights have been assigned to satisfy the mirage conditions.
In our analysis,  we have evaluated all the couplings at the running
top mass $\bar m_t \simeq 163$~GeV in the $\overline{\text{MS}}$ scheme.
We have also included the loop contributions from the sbottom
and stau adopting the results
from Refs.~\cite{Carena:1994bv, Carena:2011aa}.
The analytic expression~(\ref{Higgs-mass}) is valid if $\tan\beta$ is
moderate or large,  and if the splitting
of the stop mass eigenvalues is small~\cite{Carena:1995bx}
\begin{equation}
\frac{m^2_{\tilde t_2} - m^2_{\tilde t_1} }{m^2_{\tilde t_2} + m^2_{\tilde t_1} } \lesssim 0.5.
\end{equation}
We have confirmed that the parameter space considered in our analysis
satisfies the above condition.

\begin{figure}[tb!]
  \begin{center}
    \includegraphics[width=0.45\textwidth]{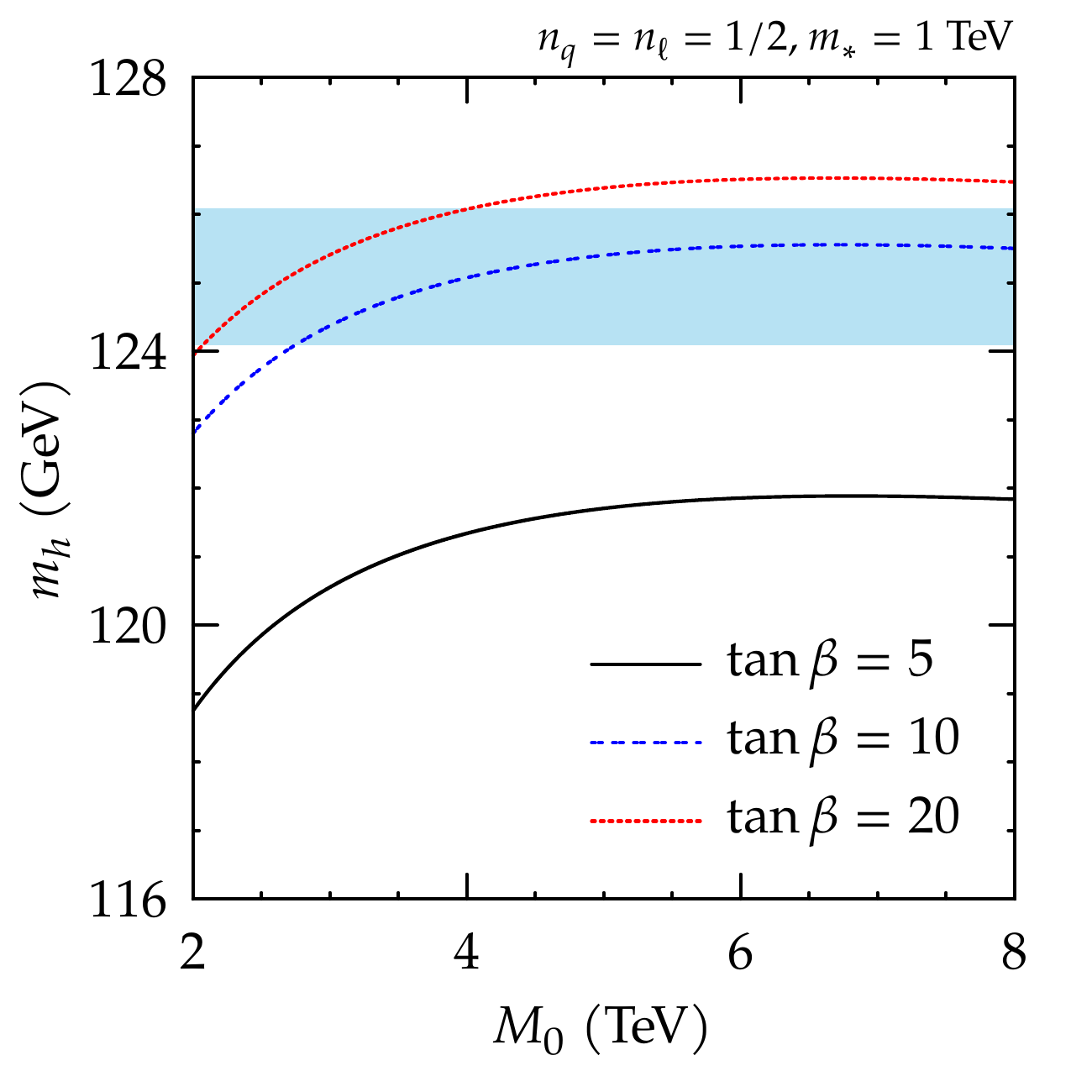}
  \end{center}
  \caption{\label{fig:mh}
    Mass of the SM-like Higgs boson as a function of the SUSY breaking scale
    $M_0$.  Here we have fixed the Higgsino mass parameter $\mu$ under the condition
    of precise gauge coupling unification by taking the SUSY threshold scale to be $m_\ast=1$~TeV.
    The blue-shaded band corresponds to $m_h = 125.09
    \pm 1$~GeV.}
\end{figure}

In Fig.~\ref{fig:mh}, we display the Higgs boson mass as a function of
the SUSY breaking scale,  $m_h(M_0)$,
for the mirage mediation with soft SUSY-breaking terms given by~(\ref{soft1}) and~(\ref{soft2})
at the mirage messenger scale $M_{\rm mir} \sim M_0$.
Note that the Higgs sector parameters,  $\mu$,  $B$, and $m^2_{H_u}$ are fixed by the unification
condition~(\ref{unification})
and the minimization conditions,  respectively.
For large $M_0$,  the stop mixing is sizable but smaller than the
maximal mixing where $|X_t| = \sqrt{6} M_S$.
Taking into account the fact that light Higgsinos are favored for
reducing the electroweak fine-tuning, we have taken the SUSY threshold
scale to be $m_\ast = 1$~TeV
to fix the value of $\mu$ by the unification condition.
We will discuss the effect of the SUSY threshold scale shortly.
Currently,  the uncertainty of the combined measurement of the Higgs
boson mass from the ATLAS and CMS experiments at the LHC is
$0.24$~GeV~\cite{Aad:2015zhl}.
However,  because the theoretical uncertainty of the Higgs mass calculation in
the SUSY models is typically about a few GeV~\cite{Athron:2016fuq,Allanach:2018fif, Bahl:2019hmm},
we show the Higgs boson mass lying in the range of $m_h =125.09 \pm
1$~GeV in the figure.
From the numerical analysis, we find that $\tan\beta$ should be large
enough to have $m_h$ around the measured value.
For $m_\ast = 1$~TeV,  the lower bound is numerically found to be
$\tan\beta \gtrsim 8$.
One can also see that a larger $M_0$ value is required for smaller $\tan\beta$.

\begin{figure}[tb!]
  \begin{center}
    \includegraphics[width=0.45\textwidth]{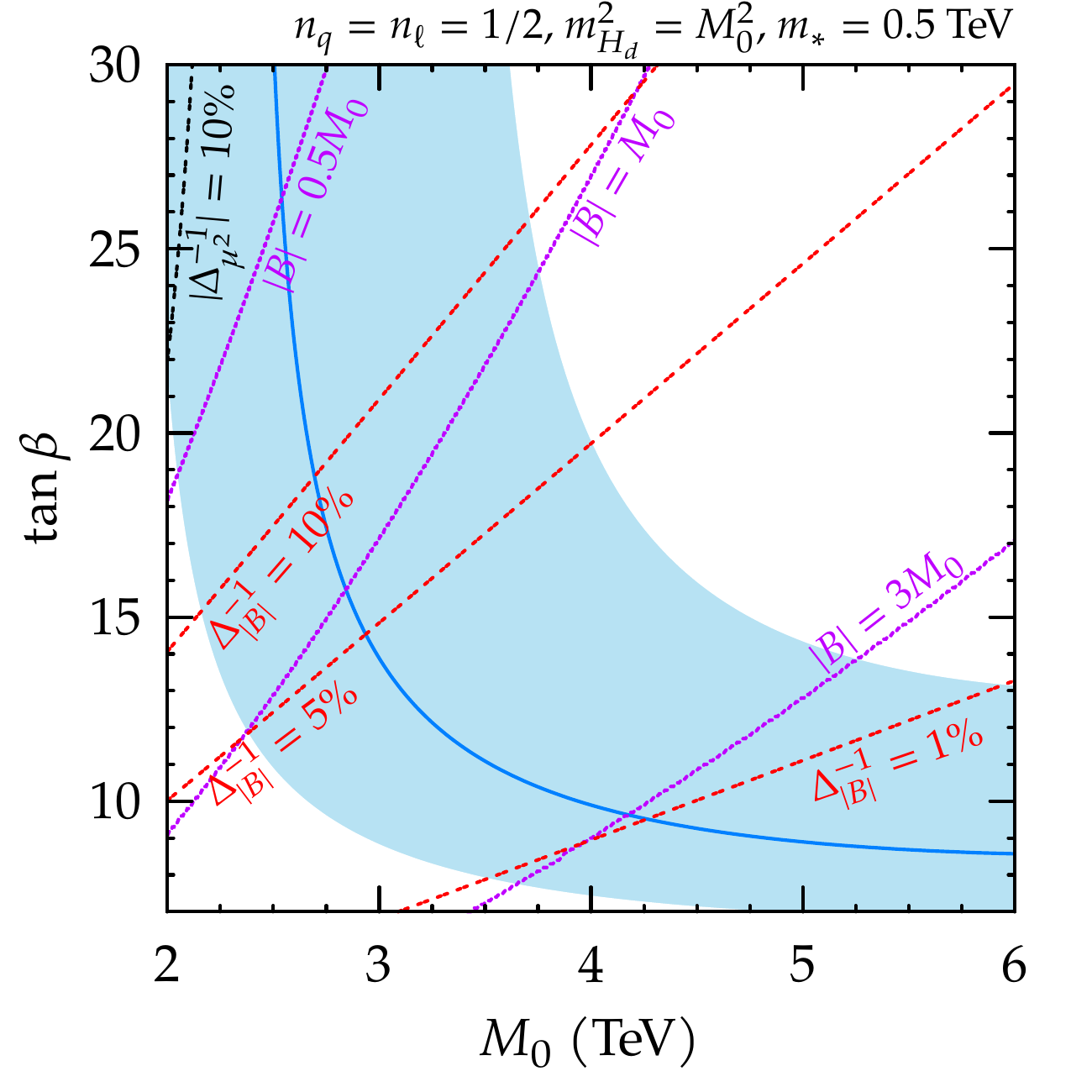}
    \includegraphics[width=0.45\textwidth]{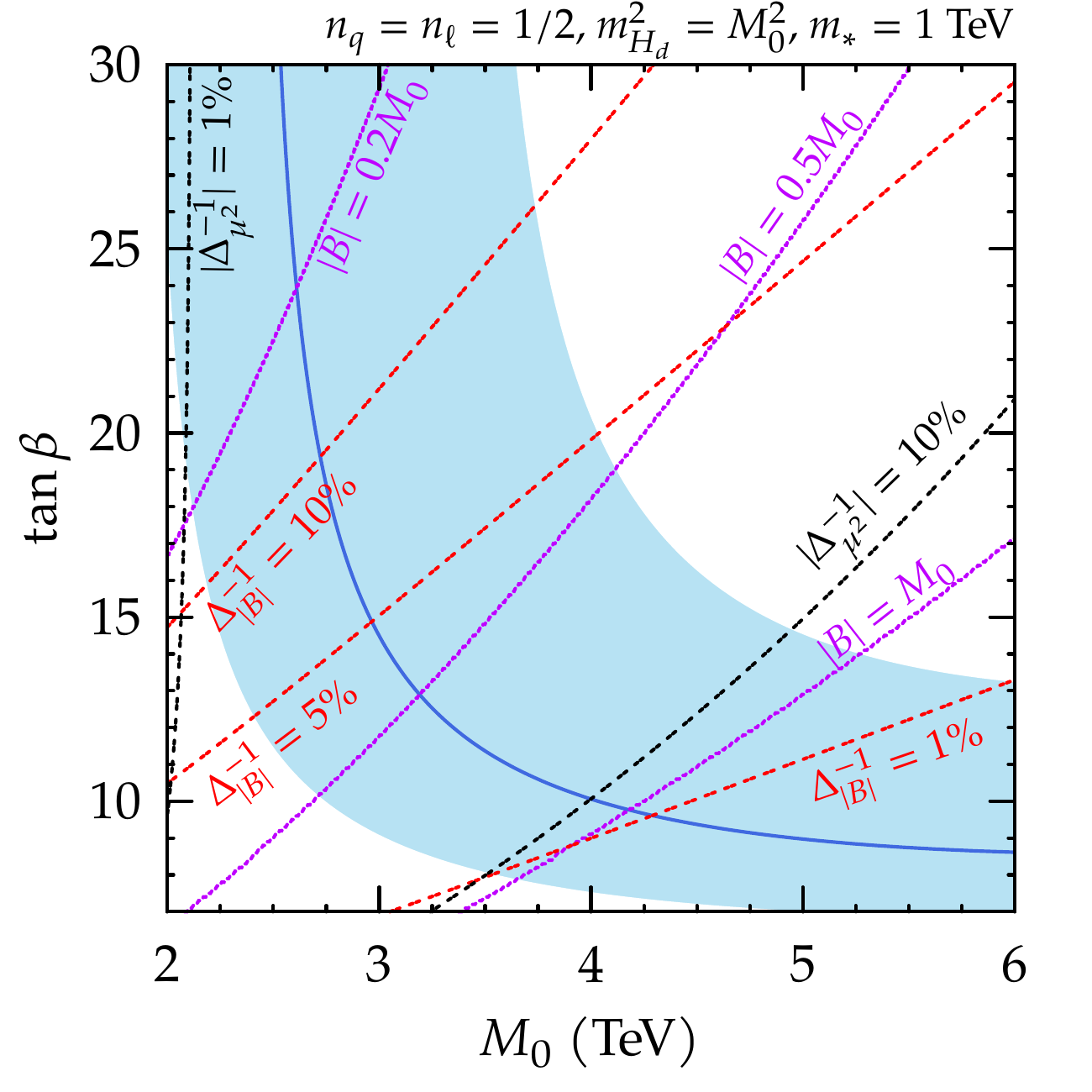}
  \end{center}
  \caption{\label{fig:m0}
    Parameter space of $M_0$ and $\tan\beta$ consistent with the
    Higgs boson mass. for $m_\ast = 0.5$~TeV (left) and  $1$~TeV (right).
    The contours of the degrees of fine-tuning for
    the EWSB are also exhibited. See the text for the details.
    The blue-shaded region corresponds to $m_h = 125.09 \pm 1$~GeV.}
\end{figure}

The values of $M_0$ and $\tan\beta$ leading to the correct Higgs
boson mass can be seen in Fig.~\ref{fig:m0}.
Again, the blue-shaded region corresponds to the parameter space of
$m_h = 125.09 \pm 1$~GeV in each panel.
It shows that the $M_0$ value tends to be larger for smaller $\tan\beta$.
For $m_h = 125$~GeV and $\tan\beta = 10$,  we need $M_0 \simeq 4$~TeV.
In the figure,  we have taken $m_\ast = 0.5$~TeV and $1$~TeV in the
left and right panels,  respectively.
The $m_\ast$ value does not affect the $M_0$ value  but it does affect model
parameters, such as $B$, by changing the Higgsino mass parameter through
the condition of gauge coupling unification as given in Eq.~(\ref{unification}).
In most of the parameter space,  we find that $|B|$ is around or below
$M_0$,  which lies in the range expected in the models as discussed in
the previous section.
%
%
For $m_\ast = 1$~TeV and $\tan\beta = 10$, $|B| \simeq M_0$.
The experimental constraint on the $M_0$ value comes from the gluino
searches at the LHC\@.
Currently, the lower bound on the gluino mass is about $2.2$~TeV~\cite{Aaboud:2017vwy, Sirunyan:2019xwh}.
In Fig.~\ref{fig:m0},  one can see that the region with large
$\tan\beta$ already starts to be excluded due to the gluino bound.
The $M_0$ parameter also receives constraints from stop
searches at the LHC~\cite{Aad:2020qwe, Sirunyan:2020tyy},
but the corresponding bound is weaker than that from the gluino searches;
$M_0 \gtrsim 1.6$~TeV for $m_{\tilde t_1} > 1$~TeV.

As the benchmark points of the TeV scale mirage mediation with $\alpha
= 2$ and $n_{H_u} = 1$, we present the Higgs and sparticle mass
spectra for the input parameters at $Q = M_{\rm mir}$ in
Table~\ref{tab:spectrum}.
The mass spectra have been calculated with \texttt{SOFTSUSY}~\cite{Allanach:2001kg}.
\begin{table}[tb!]
  \caption{\label{tab:spectrum}
    Higgs and sparticle mass spectra in GeV for (a) smaller and (b)
    larger $M_0$.
    The first and second generation sfermions are degenerate in
    mass; $m_{c_i} \simeq m_{u_i}$, $m_{s_i} \simeq
    m_{d_i}$, and $m_{\mu_i} \simeq m_{e_i}$ for $i = L$, $R$.
    The neutralino masses are taken to be positive.
    In the bottom rows, we exhibit the values of the fine-tuning measures for
    each point.
  }
  \begin{center}
  \begin{tabular}{c | c  c}
    \hline\hline
    &&\\[-2mm]
    Parameters & (a) & (b) \\[2mm]
    \hline&&\\[-2mm]
    $M_0$~(GeV) & $2800$ & $4000$ \\
    $\mu$~(GeV) & $548$ & $445$ \\
    $\tan\beta$ & $25$ & $10$ \\
    $M_{\rm mir}$~(GeV) & $700$ & $1000$ \\
    $m_{3/2}$~(TeV) & $169.65$ & $272.34$ \\[2mm]
    \hline&&\\[-2mm]
    $m_h$ & $125.1$ & $125.6$ \\
    $m_H$, $m_A$, $m_{H^+}$
    & $2855$, $2855$, $2856$
    & $4044$, $4044$, $4045$ \\
    $m_{\tilde g}$ & $2686$ & $3842$ \\
    $m_{\tilde\chi_1^0}$, $m_{\tilde\chi_2^0}$, $m_{\tilde\chi_3^0}$, $m_{\tilde\chi_4^0}$
    & $558$, $560$, $2788$, $2818$
    & $455$, $457$, $3981$, $4024$ \\
    $m_{\tilde\chi_1^+}$, $m_{\tilde\chi_2^+}$
    & $559$, $2788$
    & $456$, $3981$ \\
    $m_{\tilde t_1}$,  $m_{\tilde t_2}$
    & $1796$, $2030$
    & $2609$, $2839$ \\
    $m_{\tilde b_1}$,  $m_{\tilde b_2}$
    & $1912$, $1940$
    & $2714$, $2735$ \\
    $m_{\tilde u_L}$, $m_{\tilde u_R}$, $m_{\tilde d_L}$, $m_{\tilde d_R}$
    & $1890$, $1905$, $1891$, $1905$
    & $2704$, $2726$, $2705$, $2724$  \\
    $m_{\tilde \tau_1}$, $m_{\tilde \tau_2}$
    & $1973$, $1991$
    & $2808$, $2819$ \\
    $m_{\tilde e_L}$, $m_{\tilde e_R}$
    & $1964$, $1970$
    & $2806$, $2815$ \\[2mm]
    \hline&&\\[-2mm]
    $\Delta_{|\mu|^2}$, $\Delta_{|B|}$, $\Delta_{m_{H_d}^2}$, $\Delta_{m_{H_u}^2}$
    & $-69.7$, $6.3$, $-3.0$, $70.5$
    & $-9.5$, $80.3$, $-39.7$, $10.1$\\
    $\Delta_{M_0^2}$, $\Delta_{\kappa_{H_d}}$, $\Delta_{\kappa_{H_u}}$
    & $67.5$, $0$, $70.6$
    & $-29.6$, $0$, $10.1$ \\[2mm]
    \hline\hline
  \end{tabular}
  \end{center}
\end{table}
The soft SUSY-breaking parameters are set by the mirage relations
given in~(\ref{soft1}) and~(\ref{soft2}),
and the Higgsino mass parameter $\mu$ is fixed by the
condition of the gauge coupling unification in~(\ref{mu-gut}) with
$m_\ast = 1$~TeV.
We have taken $\mu > 0$ and $m_t = 173$~GeV.
The mirage messenger scale $M_{\rm mir}$ is determined as
in~(\ref{eq:M_mir}) for the gravitino mass $m_{3/2}$ obtained using
the relation~(\ref{eq:alpha}) and $M_{\rm GUT} = 10^{16}$~GeV.
For illustration, $M_{\rm mir}$ and $m_{3/2}$ are also shown in
Table~\ref{tab:spectrum}.
We note that in the benchmark point with smaller $M_0$, the gluino
mass is right within the reach of the high-luminosity
LHC for an integrate luminosity of 3~ab$^{-1}$~\cite{Baer:2016wkz},
while in the other benchmark point with larger $M_0$, it is far beyond
the reach.

Let us continue to discuss the fine-tuning issue. In order to estimate
quantitatively to what extent the $Z$-boson mass is sensitive to
the variations of the Higgs sector parameters, we take the
conventional fine-tuning measure defined by
\begin{equation}
\Delta_a  \equiv \frac{\partial \ln m^2_Z}{\partial \ln a },
\end{equation}
where $a$ stands for the parameters involved in the Higgs scalar
potential~\cite{Ellis:1986yg, Barbieri:1987fn,Dimopoulos:1995mi}.\footnote{
There are also other ways to quantify the degree of fine-tuning for
the electroweak symmetry breaking.
See,  for example,
Ref.~\cite{Baer:2013gva} for more discussion. We also refer the
reader to Refs.~\cite{Baer:2019tee, Baer:2020sgm} for a recent
discussion of electroweak naturalness using other convention of
the fine-tuning measure in general mirage mediation.
}
For moderate to large $\tan\beta$ values, the ratio between the Higgs VEVs is
given by $\tan\beta\simeq (\tilde m^2_{H_d} +\tilde m^2_{H_u} +
2|\mu|^2)/|B\mu|$. Then, in terms of the mass parameters in the Higgs sector, $a
= \{\mu$, $B$, $\tilde m^2_{H_d}$, $\tilde m^2_{H_u} \}$, the $Z$-boson mass is
expressed as follows:
\begin{equation}
  \label{mZ-Delta-a}
  \frac{1}{2} m^2_Z \simeq
  \frac{|B\mu|^2 \tilde m^2_{H_d}}{(
  \tilde m^2_{H_d} +\tilde m^2_{H_u} + 2|\mu|^2
  )^2}
  - \tilde m^2_{H_u} - |\mu|^2.
\end{equation}
%
In mirage mediation, the soft Higgs masses have the relation that
$|\tilde m^2_{H_u}|\lesssim \tilde m^2_{H_d}/8\pi^2$, which implies that
the minimization conditions of the Higgs scalar potential
can be satisfied only for
\begin{equation}
  |\mu|^2,\,\, |\tilde m^2_{H_u}| \ll \tilde m^2_{H_d}.
\end{equation}
From Eq.~(\ref{mZ-Delta-a}),  it is thus straightforward to see that the fine-tuning measures for
the $\mu$ and $B$ parameters are given as
\begin{eqnarray}
\label{Delta1}
\Delta_{\mu^2} &\simeq&
\frac{2|B\mu|^2}{m^2_Z \tilde m^2_{H_d}} - \frac{2|\mu|^2}{m^2_Z}
\simeq
\frac{2\tilde m^2_{H_u}}{m^2_Z},
\nonumber \\
\Delta_{|B|} &\simeq&
\frac{4 |B\mu|^2 }{m^2_Z \tilde m^2_{H_d}},
\end{eqnarray}
while those for $m^2_{H_u}$ and $m^2_{H_d}$ are
\begin{eqnarray}
\label{Delta2}
\Delta_{m^2_{H_d} } &\simeq&
-\frac{2|B\mu|^2}{m^2_Z \tilde m^2_{H_d}}
\simeq
-\frac{1}{2}\Delta_{|B|},
\nonumber \\
\Delta_{m^2_{H_u} } &\simeq&
-\frac{2 \tilde m^2_{H_u}}{m^2_Z} \simeq -\Delta_{\mu^2},
\end{eqnarray}
%
%
where the last equality in $\Delta_{\mu^2}$ follows from the observation that the minimization
conditions approximately require $|B\mu|^2 \simeq (\tilde m^2_{H_u} + |\mu|^2)\tilde m^2_{H_d}$.
Therefore,  the degree of fine tuning is determined by the larger of $|\Delta_{\mu^2}|$
and $\Delta_{|B|}$.
The $Z$-boson mass is more sensitive to the variation of the Higgs mixing parameter
$B$,
\begin{equation}
\Delta^{-1}_{|B|} \simeq 0.05 \times \left( \frac{\tan\beta}{15} \right)^2
\left( \frac{M_0}{3~{\rm TeV}} \right)^{-2},
\end{equation}
if $|\mu|$ is smaller than $3|B|/\tan\beta$,  i.e.,  if the Higgsinos are as light as
\begin{equation}
|\mu| < 346\,{\rm GeV}\times \left( \frac{\tan\beta}{15} \right)^{-1}
\left( \frac{M_0}{3~{\rm TeV}} \right),
\end{equation}
where we have used the minimization condition,  $|B\mu|\simeq \tilde m^2_{H_d}/\tan\beta$,
with $\tilde m^2_{H_d} \simeq M^2_0$.
Figure~\ref{fig:m0} shows the contours of the fine-tuning measures as well.
We find that the degree of fine-tuning is about a few percent or better
in the parameter space of
$2~\text{TeV} \lesssim M_0 \lesssim 6$~TeV and $8 \lesssim \tan\beta
\lesssim 25$, while being consistent with the measured Higgs boson mass.
Multi-TeV SUSY in the mirage mediation can therefore achieve the
electroweak symmetry breaking as naturally as the weak-scale SUSY\@.

The fine-tuning measures (\ref{Delta1}) and (\ref{Delta2}) have been obtained by varying the Higgs
sector mass couplings at the weak scale,
which are determined by the input parameters,   $M_0$,  $M_{\rm mir}$,  $n_i$,  and $\kappa_i$.
One may be concerned about the sensitivity of the $Z$-boson mass to
the variation of the input parameters and severer fine-tuning than the above
estimation.
It is, however, not the case, as one can see from
\begin{eqnarray}
\Delta_{M^2_0} &=&
\Delta_{m^2_{H_d}} + \Delta_{m^2_{H_u}},
\nonumber \\
\Delta_{\kappa_{H_d}} &\simeq&
\frac{\kappa_{H_d}}{8\pi^2  } \Delta_{m^2_{H_d}},
\nonumber \\
\Delta_{\kappa_{H_u}}  &=& \Delta_{m^2_{H_u}},
\end{eqnarray}
where we have used that  $\kappa_i$ are of the order unity,
and the modular weights are assigned by $n_{H_u}=1$ and $n_{H_d}=0$.
In Table~\ref{tab:spectrum}, we exhibit the values of $\Delta_{M_0^2}$ and
$\Delta_{\kappa_{H_d}}$, and $\Delta_{\kappa_{H_u}}$ for the benchmark points of mirage
mediation with $\alpha = 2$ and $n_{H_u} = 1$.
For the variation of $n_i$ and $M_{\rm mir}$,
we note that
the modular weights $n_i$ are not continuous but rational numbers,
$0$, $1/2$,  or $1$,
determined by the location of the corresponding matter in extra dimensions.
This implies
that the choice of modular weights is not a fine-tuning.
The $\alpha$ parameters is also a rational number, and
taking $\alpha=2$ leads to
\begin{equation}
M_{\rm mir} \simeq
 \frac{2M_{\rm GUT}}{M_{Pl}}
\ln\left(
\frac{M_{Pl}/M_0}{2\ln(M_{Pl}/M_0)}
\right)\times M_0.
\end{equation}
Here,
the unification scale is given by $M_{\rm GUT}\approx 10^{16}$~GeV,
which is insensitive
to the value of $M_0$ because it is determined by the ratios of the sparticle masses,
i.e.,~by the choice of modular weights.
The above expression shows that $M_{\rm mir}$ is fixed by the
$M_0$ value. Specifically, it is approximately proportional to
$M_0$, e.g., $M_{\rm mir}\approx 0.24 M_0$ for $M_0$ between TeV and PeV scales.
The mirage messenger scale appears in soft SUSY breaking couplings, renormalized at
the SUSY breaking scale only through
the combination of $\ln(M_{\rm mir}/M_0)$,  which rarely changes under the variation of $M_0$
for $\alpha=2$.
As is well known,
the stop sector can significantly affect the $Z$-boson mass through
the RG running and loop-potential contributions to the up-type Higgs soft mass
squared,~(\ref{mhu}) and~(\ref{mhu2}).
In mirage mediation, the contributions are proportional to $m^2_{\tilde t} \ln(M_{\rm mir}/m_{\tilde t})$
and $m^2_{\tilde t}$,  respectively,
with the stop mass given by $m_{\tilde t}\simeq \sqrt{1-n_q} M_0$.
It is therefore obvious that in the case of $\alpha = 2$,
both the stop contributions are of
the order of $M^2_0/8\pi^2$,
and they are sensitive only to the variation of $M_0$.
The effects are included in $\Delta_{M^2_0}$, or equivalently, in $\Delta_{m^2_{H_u}}$.
These observations explain why the degree of fine-tuning can be reduced up to
a loop factor, compared to the conventional scenarios.

\begin{figure}[tb!]
  \begin{center}
    \includegraphics[width=0.45\textwidth]{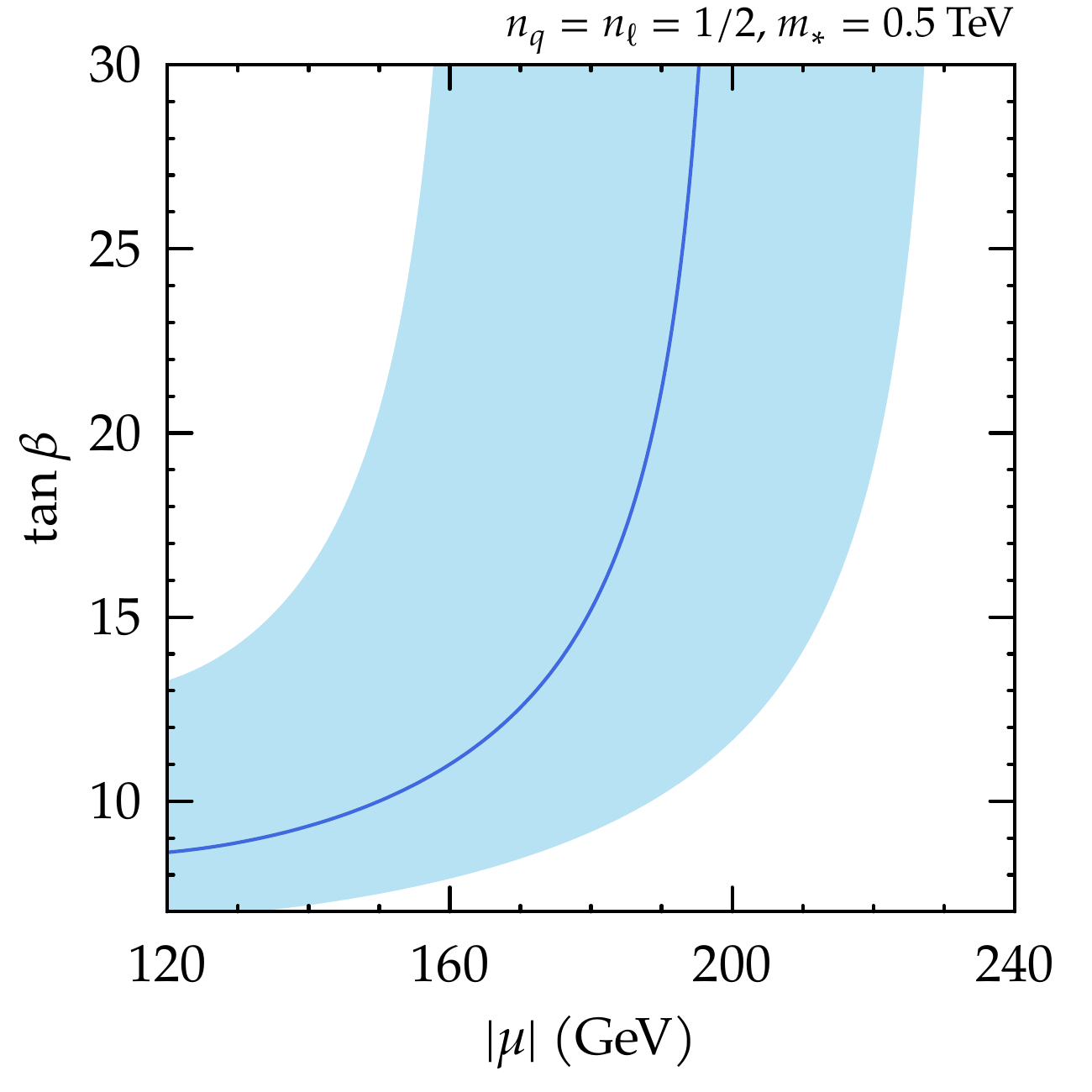}
    \includegraphics[width=0.45\textwidth]{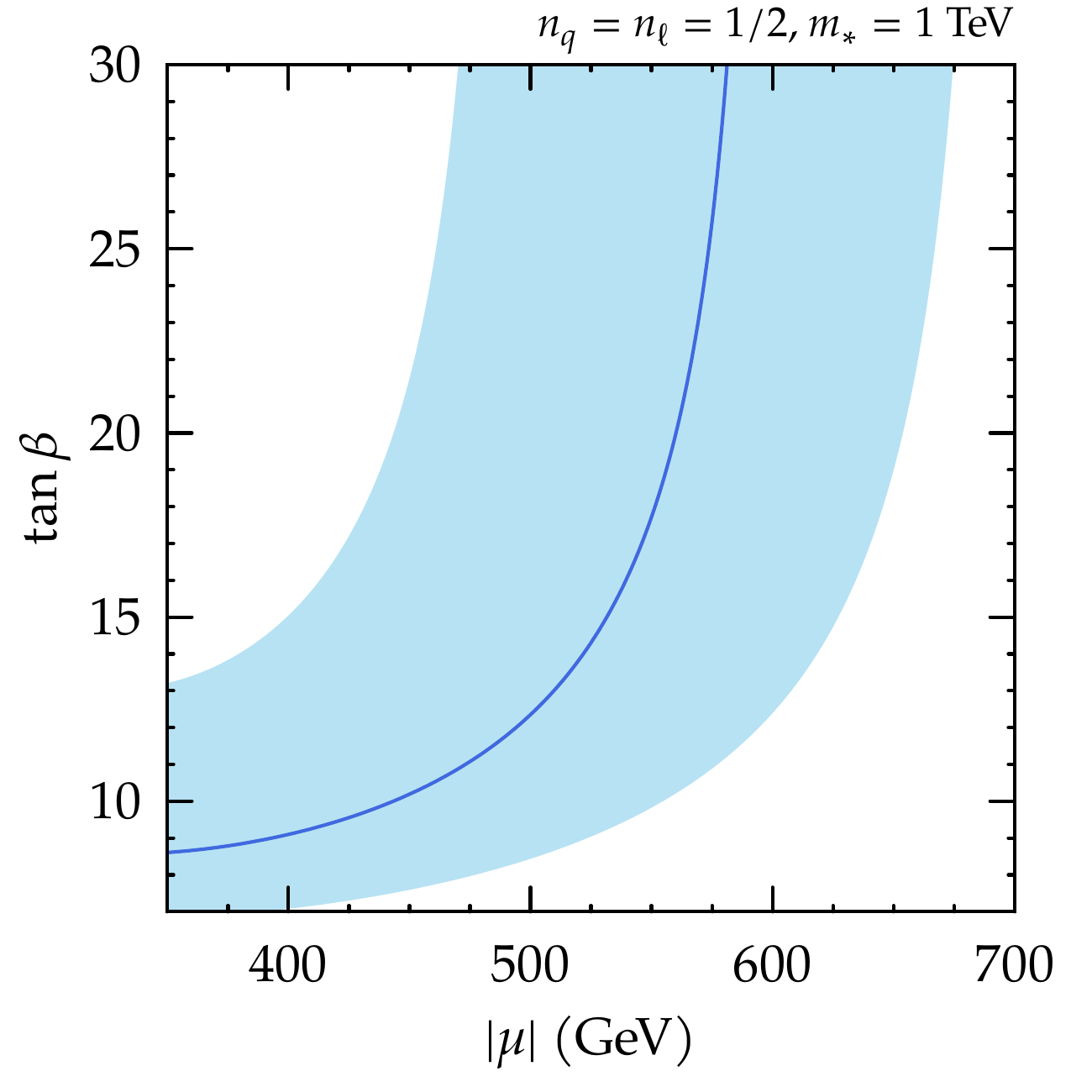}
  \end{center}
  \caption{\label{fig:mu}
    Parameter space of $\mu$ and $\tan\beta$ consistent with the Higgs boson mass
    for $m_\ast = 0.5$~TeV (left) and  $1$~TeV (right).
    The blue-shaded region corresponds to $m_h = 125.09 \pm 1$~GeV.
    }
\end{figure}

We finally discuss the effect of the SUSY threshold scale.
A precise gauge coupling unification is achieved for $m_\ast \simeq 2$~TeV,
assuming that the threshold corrections induced by heavy fields of a
grand unified theory are absent~\cite{Krippendorf:2013dqa}.
If one allows a 3\% deviation of the strong coupling~\cite{Raby:2009sf},
$m_\ast$ could be as small as $200$~GeV.
Setting $m_\ast = 1$~TeV gives rise to about 1\% deviation.
We note that the size of the $\mu$ parameter increases as $m_\ast$
becomes larger,  as can be seen in Eq.~(\ref{unification}).
Figure~\ref{fig:mu} shows the parameter space of $|\mu|$ and
$\tan\beta$ consistent with the measured Higgs mass.
One important experimental constraint concerning the $\mu$ parameter
is the lower bound on the charged Higgsino mass set by the LEP
experiment; $|\mu| > 104$~GeV.
The bound can be translated into the lower bound on $m_\ast$,  which is
$m_\ast \gtrsim 300$~GeV for $M_0 \gtrsim 2$~TeV.
Meanwhile,  a large $m_\ast \gtrsim 1.5$~TeV gives rise to $|\mu| \gtrsim
1$~TeV,  which results in too severe electroweak fine-tuning induced
by the $\mu$ parameter; $|\Delta_{\mu^2}^{-1}| \lesssim 0.01$.
Therefore,  it can be argued that $m_\ast$ is bounded from above by
the naturalness of the Higgs sector.
We conclude that the SUSY threshold scale of about $300$~GeV to a TeV
is consistent with the successful gauge coupling unification within a
deviation below a few percent while satisfying the mass bound on the
charged Higgsino and the electroweak naturalness.

\section{Conclusions}

Combined with the null results of SUSY searches at collider
experiments so far, the Higgs boson at $125$~GeV indicates that SUSY
may exist around or above the multi-TeV scale.
However,  even in such a high-scale SUSY scenario,  the electroweak
symmetry breaking can naturally occur with low fine-tuning if the
sparticles obtain masses via mirage mediation as in the KKLT flux
compactification.
It is because the up-type Higgs soft mass can be suppressed at low-energy
scales,  insensitively to the SUSY breaking scale,  and
light Higgsinos much below the SUSY breaking scale are compatible
with the gauge coupling unification.
We note that,  if the Higgsinos are around or below a few hundred GeV,
mirage mediation can serve low-electroweak fine-tuning better than a
few percent for stops between about $2$~TeV and $6$~TeV,  while
accommodating the Higgs boson consistent with the current experimental data.
We expect that the electroweak naturalness with light Higgsinos will
be tested by searching for the light neutralinos and charginos at
future colliders.

\acknowledgments

   This work was supported by IBS under Project Code No.~IBS-R018-D1 (C.B.P.),
   and by the National Research Foundation of Korea (NRF) grant funded by the Korea
   government Grant No.~2018R1C1B6006061 and No.~2021R1A4A5031460 (K.S.J.).

\appendix
\section{\label{sec:threshold_corr}Gauge threshold corrections}

In this Appendix we examine how much the mirage mediation pattern is
affected by gauge threshold corrections.
Let us suppose that the gauge couplings receive
threshold corrections at $m_{\rm th}$,   and consequently do not unify at $M_{\rm GUT}$
if extrapolated with the MSSM beta functions for running.
At energy scales below $m_{\rm th}$,   the gauge couplings are written as
\begin{equation}
\frac{1}{g^2_a(Q)}
= \frac{1}{g^2_0}
+  \frac{b_a}{4\pi^2} \ln\left( \frac{M_{\rm GUT}}{Q} \right)
+ \Delta_a,
\end{equation}
where
$\Delta_a$ is the threshold correction to the gauge coupling, and
the universal part is determined by the gauge kinetic function,
$g^2_0=1/{\rm Re}(f_a)$,  with $f_a = T + {\rm constant}$.
Using the modulus dependence of gauge couplings,   one finds the gaugino masses to be
\begin{equation}
M_a(m_{\rm th}) =  \frac{g^2_a(m_{\rm th})}{g^2_0} M_0
+ \frac{b_a g^2_a(m_{\rm th})}{8\pi^2} m_{3/2},
\end{equation}
at the scale just below $m_{\rm th}$.
Here the term proportional to $b_a$ is the anomaly-mediated contribution,  which
does not depend on physics at higher-energy scales.
Because the combination $M_a/g^2_a$ does not run with the scale,
low-energy gaugino masses are obtained by
\begin{equation}
M_a(Q) =
M_{a0}
\left\{
1 - \frac{b_a g^2_a(Q)}{4\pi^2}\ln\left( \frac{ M_{a{\rm mir} } }{Q} \right)  \right\},
\end{equation}
for $M_{a0}$ and $M_{a{\rm mir}}$ given by
\begin{align}
M_{a0} &=  \frac{M_0}{1+\epsilon_a},
\nonumber \\
M_{a{\rm mir}} &=
M_{\rm GUT} \left( \frac{m_{3/2}}{M_{Pl}} \right)^{\frac{\alpha}{2} (1+\epsilon_a)},
\end{align}
where $\epsilon_a \equiv  g^2_0 \Delta_a$ parametrize the threshold corrections.
The above shows that the unification of gaugino masses does
not occur if the threshold corrections are nonuniversal.
For universal $\epsilon_a$,  the gauginos have a common mass at the
mirage-messenger scale,   but its value is different from $M_0$ unless
$\epsilon_a$ vanish.\footnote{
Deflected mirage mediation~\cite{Everett:2008qy, Everett:2008ey, Choi:2009jn}
is an example of universal gauge-threshold corrections.
}
It is only when $\epsilon_a=0$ that mirage mediation effectively
corresponds to  pure moduli mediation transmitted at the mirage-messenger scale.
Such correspondence is essential for reducing the electroweak
fine-tuning because it allows to suppress the up-type Higgs soft mass
at the mirage messenger scale by taking an appropriate modular weight.
We thus require that $\epsilon_a$ be at most around the loop factor,  $1/8\pi^2$, in size if any.


\bibliographystyle{JHEP}
\bibliography{high-scale-mirage}

\providecommand{\href}[2]{#2}\begingroup\raggedright\begin{thebibliography}{10}

\bibitem{Nilles:1983ge}
H.P.~Nilles, \emph{{Supersymmetry, Supergravity and Particle Physics}},
  \href{https://doi.org/10.1016/0370-1573(84)90008-5}{\emph{Phys. Rept.}
  {\bfseries 110} (1984) 1}.

\bibitem{Haber:1984rc}
H.E.~Haber and G.L.~Kane, \emph{{The Search for Supersymmetry: Probing Physics
  Beyond the Standard Model}},
  \href{https://doi.org/10.1016/0370-1573(85)90051-1}{\emph{Phys. Rept.}
  {\bfseries 117} (1985) 75}.

\bibitem{Graham:2015cka}
P.W.~Graham, D.E.~Kaplan and S.~Rajendran, \emph{{Cosmological Relaxation of
  the Electroweak Scale}},
  \href{https://doi.org/10.1103/PhysRevLett.115.221801}{\emph{Phys. Rev. Lett.}
  {\bfseries 115} (2015) 221801}
  [\href{https://arxiv.org/abs/1504.07551}{{\ttfamily 1504.07551}}].

\bibitem{Choi:2005uz}
K.~Choi, K.S.~Jeong and K.-i.~Okumura, \emph{{Phenomenology of mixed
  modulus-anomaly mediation in fluxed string compactifications and brane
  models}}, \href{https://doi.org/10.1088/1126-6708/2005/09/039}{\emph{JHEP}
  {\bfseries 09} (2005) 039}
  [\href{https://arxiv.org/abs/hep-ph/0504037}{{\ttfamily hep-ph/0504037}}].

\bibitem{Kachru:2003aw}
S.~Kachru, R.~Kallosh, A.D.~Linde and S.P.~Trivedi, \emph{{De Sitter vacua in
  string theory}},
  \href{https://doi.org/10.1103/PhysRevD.68.046005}{\emph{Phys. Rev. D}
  {\bfseries 68} (2003) 046005}
  [\href{https://arxiv.org/abs/hep-th/0301240}{{\ttfamily hep-th/0301240}}].

\bibitem{Choi:2005ge}
K.~Choi, A.~Falkowski, H.P.~Nilles and M.~Olechowski, \emph{{Soft supersymmetry
  breaking in KKLT flux compactification}},
  \href{https://doi.org/10.1016/j.nuclphysb.2005.04.032}{\emph{Nucl. Phys. B}
  {\bfseries 718} (2005) 113}
  [\href{https://arxiv.org/abs/hep-th/0503216}{{\ttfamily hep-th/0503216}}].

\bibitem{Choi:2005hd}
K.~Choi, K.S.~Jeong, T.~Kobayashi and K.-i.~Okumura, \emph{{Little SUSY
  hierarchy in mixed modulus-anomaly mediation}},
  \href{https://doi.org/10.1016/j.physletb.2005.11.078}{\emph{Phys. Lett. B}
  {\bfseries 633} (2006) 355}
  [\href{https://arxiv.org/abs/hep-ph/0508029}{{\ttfamily hep-ph/0508029}}].

\bibitem{Choi:2006xb}
K.~Choi, K.S.~Jeong, T.~Kobayashi and K.-i.~Okumura, \emph{{TeV Scale Mirage
  Mediation and Natural Little SUSY Hierarchy}},
  \href{https://doi.org/10.1103/PhysRevD.75.095012}{\emph{Phys. Rev. D}
  {\bfseries 75} (2007) 095012}
  [\href{https://arxiv.org/abs/hep-ph/0612258}{{\ttfamily hep-ph/0612258}}].

\bibitem{Choi:2006bh}
K.~Choi and K.S.~Jeong, \emph{{Supersymmetry breaking and moduli stabilization
  with anomalous U(1) gauge symmetry}},
  \href{https://doi.org/10.1088/1126-6708/2006/08/007}{\emph{JHEP} {\bfseries
  08} (2006) 007} [\href{https://arxiv.org/abs/hep-th/0605108}{{\ttfamily
  hep-th/0605108}}].

\bibitem{Kawamura:2017qey}
J.~Kawamura and Y.~Omura, \emph{{Analysis of the TeV-scale mirage mediation
  with heavy superparticles}},
  \href{https://doi.org/10.1007/JHEP11(2017)189}{\emph{JHEP} {\bfseries 11}
  (2017) 189} [\href{https://arxiv.org/abs/1710.03412}{{\ttfamily
  1710.03412}}].

\bibitem{Krippendorf:2013dqa}
S.~Krippendorf, H.P.~Nilles, M.~Ratz and M.W.~Winkler, \emph{{Hidden SUSY from
  precision gauge unification}},
  \href{https://doi.org/10.1103/PhysRevD.88.035022}{\emph{Phys. Rev. D}
  {\bfseries 88} (2013) 035022}
  [\href{https://arxiv.org/abs/1306.0574}{{\ttfamily 1306.0574}}].

\bibitem{Jeong:2017hgz}
K.S.~Jeong, \emph{{Light Higgsino for Gauge Coupling Unification}},
  \href{https://doi.org/10.1016/j.physletb.2017.03.028}{\emph{Phys. Lett. B}
  {\bfseries 769} (2017) 42}
  [\href{https://arxiv.org/abs/1701.06947}{{\ttfamily 1701.06947}}].

\bibitem{Ellis:2017erg}
S.A.R.~Ellis and J.D.~Wells, \emph{{High-scale supersymmetry, the Higgs boson
  mass, and gauge unification}},
  \href{https://doi.org/10.1103/PhysRevD.96.055024}{\emph{Phys. Rev. D}
  {\bfseries 96} (2017) 055024}
  [\href{https://arxiv.org/abs/1706.00013}{{\ttfamily 1706.00013}}].

\bibitem{Thomas:1998wy}
S.D.~Thomas and J.D.~Wells, \emph{{Phenomenology of Massive Vectorlike Doublet
  Leptons}}, \href{https://doi.org/10.1103/PhysRevLett.81.34}{\emph{Phys. Rev.
  Lett.} {\bfseries 81} (1998) 34}
  [\href{https://arxiv.org/abs/hep-ph/9804359}{{\ttfamily hep-ph/9804359}}].

\bibitem{Baer:2011ec}
H.~Baer, V.~Barger and P.~Huang, \emph{{Hidden SUSY at the LHC: the light
  higgsino-world scenario and the role of a lepton collider}},
  \href{https://doi.org/10.1007/JHEP11(2011)031}{\emph{JHEP} {\bfseries 11}
  (2011) 031} [\href{https://arxiv.org/abs/1107.5581}{{\ttfamily 1107.5581}}].

\bibitem{Berggren:2013vfa}
M.~Berggren, F.~Br\"ummer, J.~List, G.~Moortgat-Pick, T.~Robens, K.~Rolbiecki
  et~al., \emph{{Tackling light higgsinos at the ILC}},
  \href{https://doi.org/10.1140/epjc/s10052-013-2660-y}{\emph{Eur. Phys. J. C}
  {\bfseries 73} (2013) 2660}
  [\href{https://arxiv.org/abs/1307.3566}{{\ttfamily 1307.3566}}].

\bibitem{Nakamura:2008ey}
S.~Nakamura, K.-i.~Okumura and M.~Yamaguchi, \emph{{Axionic Mirage Mediation}},
  \href{https://doi.org/10.1103/PhysRevD.77.115027}{\emph{Phys. Rev. D}
  {\bfseries 77} (2008) 115027}
  [\href{https://arxiv.org/abs/0803.3725}{{\ttfamily 0803.3725}}].

\bibitem{Choi:2009qd}
K.~Choi, K.S.~Jeong, W.-I.~Park and C.S.~Shin, \emph{{Thermal inflation and
  baryogenesis in heavy gravitino scenario}},
  \href{https://doi.org/10.1088/1475-7516/2009/11/018}{\emph{JCAP} {\bfseries
  11} (2009) 018} [\href{https://arxiv.org/abs/0908.2154}{{\ttfamily
  0908.2154}}].

\bibitem{Baer:2012cf}
H.~Baer, V.~Barger, P.~Huang, D.~Mickelson, A.~Mustafayev and X.~Tata,
  \emph{{Radiative natural supersymmetry: Reconciling electroweak fine-tuning
  and the Higgs boson mass}},
  \href{https://doi.org/10.1103/PhysRevD.87.115028}{\emph{Phys. Rev. D}
  {\bfseries 87} (2013) 115028}
  [\href{https://arxiv.org/abs/1212.2655}{{\ttfamily 1212.2655}}].

\bibitem{Carena:1995wu}
M.~Carena, M.~Quiros and C.~Wagner, \emph{{Effective potential methods and the
  Higgs mass spectrum in the MSSM}},
  \href{https://doi.org/10.1016/0550-3213(95)00665-6}{\emph{Nucl. Phys. B}
  {\bfseries 461} (1996) 407}
  [\href{https://arxiv.org/abs/hep-ph/9508343}{{\ttfamily hep-ph/9508343}}].

\bibitem{Haber:1996fp}
H.E.~Haber, R.~Hempfling and A.H.~Hoang, \emph{{Approximating the radiatively
  corrected Higgs mass in the minimal supersymmetric model}},
  \href{https://doi.org/10.1007/s002880050498}{\emph{Z. Phys. C} {\bfseries 75}
  (1997) 539} [\href{https://arxiv.org/abs/hep-ph/9609331}{{\ttfamily
  hep-ph/9609331}}].

\bibitem{Carena:2000dp}
M.~Carena, H.~Haber, S.~Heinemeyer, W.~Hollik, C.~Wagner and G.~Weiglein,
  \emph{{Reconciling the two loop diagrammatic and effective field theory
  computations of the mass of the lightest CP-even Higgs boson in the MSSM}},
  \href{https://doi.org/10.1016/S0550-3213(00)00212-1}{\emph{Nucl. Phys. B}
  {\bfseries 580} (2000) 29}
  [\href{https://arxiv.org/abs/hep-ph/0001002}{{\ttfamily hep-ph/0001002}}].

\bibitem{Carena:1994bv}
M.~Carena, M.~Olechowski, S.~Pokorski and C.~Wagner, \emph{{Electroweak
  symmetry breaking and bottom - top Yukawa unification}},
  \href{https://doi.org/10.1016/0550-3213(94)90313-1}{\emph{Nucl. Phys. B}
  {\bfseries 426} (1994) 269}
  [\href{https://arxiv.org/abs/hep-ph/9402253}{{\ttfamily hep-ph/9402253}}].

\bibitem{Carena:2011aa}
M.~Carena, S.~Gori, N.R.~Shah and C.E.~Wagner, \emph{{A 125 GeV SM-like Higgs
  in the MSSM and the $\gamma \gamma$ rate}},
  \href{https://doi.org/10.1007/JHEP03(2012)014}{\emph{JHEP} {\bfseries 03}
  (2012) 014} [\href{https://arxiv.org/abs/1112.3336}{{\ttfamily 1112.3336}}].

\bibitem{Carena:1995bx}
M.~Carena, J.~Espinosa, M.~Quiros and C.~Wagner, \emph{{Analytical expressions
  for radiatively corrected Higgs masses and couplings in the MSSM}},
  \href{https://doi.org/10.1016/0370-2693(95)00694-G}{\emph{Phys. Lett. B}
  {\bfseries 355} (1995) 209}
  [\href{https://arxiv.org/abs/hep-ph/9504316}{{\ttfamily hep-ph/9504316}}].

\bibitem{Aad:2015zhl}
{\scshape ATLAS, CMS} collaboration, \emph{{Combined Measurement of the Higgs
  Boson Mass in $pp$ Collisions at $\sqrt{s}=7$ and 8 TeV with the ATLAS and
  CMS Experiments}},
  \href{https://doi.org/10.1103/PhysRevLett.114.191803}{\emph{Phys. Rev. Lett.}
  {\bfseries 114} (2015) 191803}
  [\href{https://arxiv.org/abs/1503.07589}{{\ttfamily 1503.07589}}].

\bibitem{Athron:2016fuq}
P.~Athron, J.-h.~Park, T.~Steudtner, D.~St\"ockinger and A.~Voigt,
  \emph{{Precise Higgs mass calculations in (non-)minimal supersymmetry at both
  high and low scales}},
  \href{https://doi.org/10.1007/JHEP01(2017)079}{\emph{JHEP} {\bfseries 01}
  (2017) 079} [\href{https://arxiv.org/abs/1609.00371}{{\ttfamily
  1609.00371}}].

\bibitem{Allanach:2018fif}
B.~Allanach and A.~Voigt, \emph{{Uncertainties in the Lightest $CP$ Even Higgs
  Boson Mass Prediction in the Minimal Supersymmetric Standard Model: Fixed
  Order Versus Effective Field Theory Prediction}},
  \href{https://doi.org/10.1140/epjc/s10052-018-6046-z}{\emph{Eur. Phys. J. C}
  {\bfseries 78} (2018) 573}
  [\href{https://arxiv.org/abs/1804.09410}{{\ttfamily 1804.09410}}].

\bibitem{Bahl:2019hmm}
H.~Bahl, S.~Heinemeyer, W.~Hollik and G.~Weiglein, \emph{{Theoretical
  uncertainties in the MSSM Higgs boson mass calculation}},
  \href{https://doi.org/10.1140/epjc/s10052-020-8079-3}{\emph{Eur. Phys. J. C}
  {\bfseries 80} (2020) 497}
  [\href{https://arxiv.org/abs/1912.04199}{{\ttfamily 1912.04199}}].

\bibitem{Aaboud:2017vwy}
{\scshape ATLAS} collaboration, \emph{{Search for squarks and gluinos in final
  states with jets and missing transverse momentum using 36 fb$^{-1}$ of
  $\sqrt{s}=13$ TeV pp collision data with the ATLAS detector}},
  \href{https://doi.org/10.1103/PhysRevD.97.112001}{\emph{Phys. Rev. D}
  {\bfseries 97} (2018) 112001}
  [\href{https://arxiv.org/abs/1712.02332}{{\ttfamily 1712.02332}}].

\bibitem{Sirunyan:2019xwh}
{\scshape CMS} collaboration, \emph{{Searches for physics beyond the standard
  model with the $M_\mathrm{T2}$ variable in hadronic final states with and
  without disappearing tracks in proton-proton collisions at $\sqrt{s}=$ 13
  TeV}}, \href{https://doi.org/10.1140/epjc/s10052-019-7493-x}{\emph{Eur. Phys.
  J. C} {\bfseries 80} (2020) 3}
  [\href{https://arxiv.org/abs/1909.03460}{{\ttfamily 1909.03460}}].

\bibitem{Aad:2020qwe}
{\scshape ATLAS} collaboration, \emph{{Search for top squarks in events with a
  Higgs or $Z$ boson using 139 fb$^{-1}$ of $pp$ collision data at
  $\sqrt{s}=13$ TeV with the ATLAS detector}},
  \href{https://doi.org/10.1140/epjc/s10052-020-08469-8}{\emph{Eur. Phys. J. C}
  {\bfseries 80} (2020) 1080}
  [\href{https://arxiv.org/abs/2006.05880}{{\ttfamily 2006.05880}}].

\bibitem{Sirunyan:2020tyy}
{\scshape CMS} collaboration, \emph{{Search for top squark pair production
  using dilepton final states in ${\text {p}}{\text {p}}$ collision data
  collected at $\sqrt{s}=13\,\text {TeV} $}},
  \href{https://doi.org/10.1140/epjc/s10052-020-08701-5}{\emph{Eur. Phys. J. C}
  {\bfseries 81} (2021) 3} [\href{https://arxiv.org/abs/2008.05936}{{\ttfamily
  2008.05936}}].

\bibitem{Allanach:2001kg}
B.C.~Allanach, \emph{{SOFTSUSY: a program for calculating supersymmetric
  spectra}}, \href{https://doi.org/10.1016/S0010-4655(01)00460-X}{\emph{Comput.
  Phys. Commun.} {\bfseries 143} (2002) 305}
  [\href{https://arxiv.org/abs/hep-ph/0104145}{{\ttfamily hep-ph/0104145}}].

\bibitem{Baer:2016wkz}
H.~Baer, V.~Barger, J.S.~Gainer, P.~Huang, M.~Savoy, D.~Sengupta et~al.,
  \emph{{Gluino reach and mass extraction at the LHC in radiatively-driven
  natural SUSY}},
  \href{https://doi.org/10.1140/epjc/s10052-017-5067-3}{\emph{Eur. Phys. J. C}
  {\bfseries 77} (2017) 499}
  [\href{https://arxiv.org/abs/1612.00795}{{\ttfamily 1612.00795}}].

\bibitem{Ellis:1986yg}
J.R.~Ellis, K.~Enqvist, D.V.~Nanopoulos and F.~Zwirner, \emph{{Observables in
  Low-Energy Superstring Models}},
  \href{https://doi.org/10.1142/S0217732386000105}{\emph{Mod. Phys. Lett. A}
  {\bfseries 01} (1986) 57}.

\bibitem{Barbieri:1987fn}
R.~Barbieri and G.~Giudice, \emph{{Upper Bounds on Supersymmetric Particle
  Masses}}, \href{https://doi.org/10.1016/0550-3213(88)90171-X}{\emph{Nucl.
  Phys. B} {\bfseries 306} (1988) 63}.

\bibitem{Dimopoulos:1995mi}
S.~Dimopoulos and G.~Giudice, \emph{{Naturalness constraints in supersymmetric
  theories with nonuniversal soft terms}},
  \href{https://doi.org/10.1016/0370-2693(95)00961-J}{\emph{Phys. Lett. B}
  {\bfseries 357} (1995) 573}
  [\href{https://arxiv.org/abs/hep-ph/9507282}{{\ttfamily hep-ph/9507282}}].

\bibitem{Baer:2013gva}
H.~Baer, V.~Barger and D.~Mickelson, \emph{{How conventional measures
  overestimate electroweak fine-tuning in supersymmetric theory}},
  \href{https://doi.org/10.1103/PhysRevD.88.095013}{\emph{Phys. Rev. D}
  {\bfseries 88} (2013) 095013}
  [\href{https://arxiv.org/abs/1309.2984}{{\ttfamily 1309.2984}}].

\bibitem{Baer:2019tee}
H.~Baer, V.~Barger and D.~Sengupta, \emph{{Mirage mediation from the
  landscape}},
  \href{https://doi.org/10.1103/PhysRevResearch.2.013346}{\emph{Phys. Rev.
  Res.} {\bfseries 2} (2020) 013346}
  [\href{https://arxiv.org/abs/1912.01672}{{\ttfamily 1912.01672}}].

\bibitem{Baer:2020sgm}
H.~Baer, V.~Barger, S.~Salam, D.~Sengupta and X.~Tata, \emph{{The LHC higgsino
  discovery plane for present and future SUSY searches}},
  \href{https://doi.org/10.1016/j.physletb.2020.135777}{\emph{Phys. Lett. B}
  {\bfseries 810} (2020) 135777}
  [\href{https://arxiv.org/abs/2007.09252}{{\ttfamily 2007.09252}}].

\bibitem{Raby:2009sf}
S.~Raby, M.~Ratz and K.~Schmidt-Hoberg, \emph{{Precision gauge unification in
  the MSSM}}, \href{https://doi.org/10.1016/j.physletb.2010.03.060}{\emph{Phys.
  Lett. B} {\bfseries 687} (2010) 342}
  [\href{https://arxiv.org/abs/0911.4249}{{\ttfamily 0911.4249}}].

\bibitem{Everett:2008qy}
L.L.~Everett, I.-W.~Kim, P.~Ouyang and K.M.~Zurek, \emph{{Deflected Mirage
  Mediation: A Framework for Generalized Supersymmetry Breaking}},
  \href{https://doi.org/10.1103/PhysRevLett.101.101803}{\emph{Phys. Rev. Lett.}
  {\bfseries 101} (2008) 101803}
  [\href{https://arxiv.org/abs/0804.0592}{{\ttfamily 0804.0592}}].

\bibitem{Everett:2008ey}
L.L.~Everett, I.-W.~Kim, P.~Ouyang and K.M.~Zurek, \emph{{Moduli Stabilization
  and Supersymmetry Breaking in Deflected Mirage Mediation}},
  \href{https://doi.org/10.1088/1126-6708/2008/08/102}{\emph{JHEP} {\bfseries
  08} (2008) 102} [\href{https://arxiv.org/abs/0806.2330}{{\ttfamily
  0806.2330}}].

\bibitem{Choi:2009jn}
K.~Choi, K.S.~Jeong, S.~Nakamura, K.-I.~Okumura and M.~Yamaguchi,
  \emph{{Sparticle masses in deflected mirage mediation}},
  \href{https://doi.org/10.1088/1126-6708/2009/04/107}{\emph{JHEP} {\bfseries
  04} (2009) 107} [\href{https://arxiv.org/abs/0901.0052}{{\ttfamily
  0901.0052}}].

\end{thebibliography}\endgroup

\end{document}